\documentclass[a4paper, 11pt]{article}

\usepackage{graphicx}
\usepackage{multirow}
\usepackage{amsmath,amssymb,amsfonts}
\usepackage{amsthm}
\usepackage{mathrsfs}
\usepackage[title]{appendix}
\usepackage{xcolor}
\usepackage{textcomp}
\usepackage{manyfoot}
\usepackage{booktabs}
\usepackage{algorithm}
\usepackage{algorithmicx}
\usepackage{algpseudocode}
\usepackage{geometry}
\usepackage[colorlinks=true]{hyperref}

\colorlet{mycolor}{black!20!blue!80!}
\hypersetup{urlcolor=blue, citecolor=blue}

\usepackage{listings}%

\usepackage[compress]{cite}

\newtheorem{Theorem}{Theorem}

\newcommand{\myalphfoot}
{
	\renewcommand{\thefootnote}{\alph{footnote}}
}

\begin{document}

\title{Applying Changes to  LDPC Codes by Using 3-Induced Matchings and Eigenvalues}

\myalphfoot

\author{
	\myalphfoot
	Alireza Eshraghi Dehaghani\footnotemark[1] , Sara Saeedi Madani\footnotemark[2] , Dariush Kiani\footnotemark[3]}

\footnotetext[1]{Department of Mathematics and Computer Science, Amirkabir University of Technology (Tehran Polytechnic), Tehran, Iran (e-mail: a\_eshraghi@aut.ac.ir).}
\footnotetext[2]{Department of Mathematics and Computer Science, Amirkabir University of Technology (Tehran Polytechnic), Tehran, Iran, and the School of Mathematics,
Institute for Research in Fundamental Sciences (IPM), P.O. Box: 19395-5746, Tehran, Iran (e-mail: sarasaeedi@aut.ac.ir).} 
\footnotetext[3]{Department of Mathematics and Computer Science, Amirkabir University of Technology (Tehran Polytechnic), Tehran, Iran (e-mail: dkiani@aut.ac.ir).}

\date{}

\maketitle

\begin{abstract}
The 2-switch operation for reducing the number of short cycles in Tanner graphs of LDPC codes has been utilized in some works. If this operation is appropriately applied  to a 2-induced matching in a bipartite graph, then it surely preserves the bipartiteness of the graph.  With a 2-induced matching, we have only one choice for applying this type of change to the graph. Using 3-induced matchings creates more potential and flexibility to apply  certain changes to the Tanner graph. We present a necessary and sufficient condition  for the existence of a  3-induced matching in bipartite graphs with girth at least $6$.
A 3-induced matching helps us to have more choices for applying a change to a bipartite graph while surely preserving the bipartiteness of the graph and keeping  the degree sequences unchanged. Among all changes that we consider, 
we select the best change  for reducing the number of cycles of length 6.
The eigenvalues of subgraphs are used  to reach a 3-induced matching, and a condition in some steps can be checked to ensure that the subgraph has at least one 3-induced matching. 
We present some numerical results regarding the steps at which the condition that guarantees the existence of a 3-induced matching is checked.
\end{abstract}


\section{Introduction}

One important factor affecting the performance of LDPC codes is the number of short cycles in Tanner graphs  \cite{Halford}. In multiple studies, attempts have been made to construct Tanner graphs with  good  cycle properties  with the help of different tools; see, for example, \cite{Xu 1,Xu 2,Jiang 1,Kim 1}. Applying changes to the Tanner graph so that the degree sequences of the graph remain constant and the number of short  cycles is reduced is a valuable purpose in coding. In  this context, in \cite{Dehghan 1},  an operation  called \emph{2-switch} has been used for removing  short cycles in bipartite graphs.
Regarding  applying a 2-switch  to  bipartite graphs, 2-induced matchings have special importance because they allow the operation to be done without adding any edge whose endpoints lie within the same partition of the graph.
In \cite{Eshraghi 1}, some necessary and sufficient conditions for the existence of a 2-induced matching in a bipartite graph have been presented in terms of the eigenvalues and degree sequences, and also a process has been introduced based on the  aforementioned conditions and  considering the smallest gap between the two largest eigenvalues ​​of the graph.

In this work, we discuss utilizing  3-induced matchings and several strategies based on the eigenvalues of vertex-induced subgraphs. A 3-induced matching has more flexibility for applying changes to graphs compared to a 2-induced matching. We use the eigenvalues of subgraphs (similar to \cite{Eshraghi 1}) to create a sequence of nested vertex-induced subgraphs to finally identify the  appropriate vertices in a graph.  On the other hand, in some steps, we use an important condition to ensure that the subgraph has a 3-induced matching. In fact, we want to find six special vertices in the graph and  test the effect of five changes in the graph  separately. Among the five new graphs that are obtained, the graph  whose girth is at least  $6$ and has the least number of cycles of length $6$ is considered as our new graph, and this process can continue in this way (more details are presented in Section \ref{section 4}).
We also present some curves illustrating the relationship between the steps where the main  condition mentioned above is checked and the achievement of a 3-induced matching in the final step.

\section{Preliminaries}\label{section 2}

Let $G$ be an undirected simple  graph. We denote
the set of vertices and the set of edges of $G$ by $V(G)$ and $E(G)$, respectively. Also, we set $p(G)=|V(G)|$ and $q(G)=|E(G)|$. We denote the set of neighbors of a vertex $v$ by $N(v)$ and its degree  by $d(v)$. 
If $v_1, \dots , v_m \in V(G)$, then we define $N(v_1, \dots , v_m)$  as 
$
N(v_1)\cup \dots \cup N(v_m)
$.
Also, we denote a complete graph with $n$ vertices by $K_n$. We say that the edges $e_{1},\dots, e_k$ are independent if  they are pairwise disjoint. 
A matching with $i$  edges is a set of $i$ independent edges.

\begin{figure}[h!]
	\begin{center}
		\includegraphics[scale=0.47]{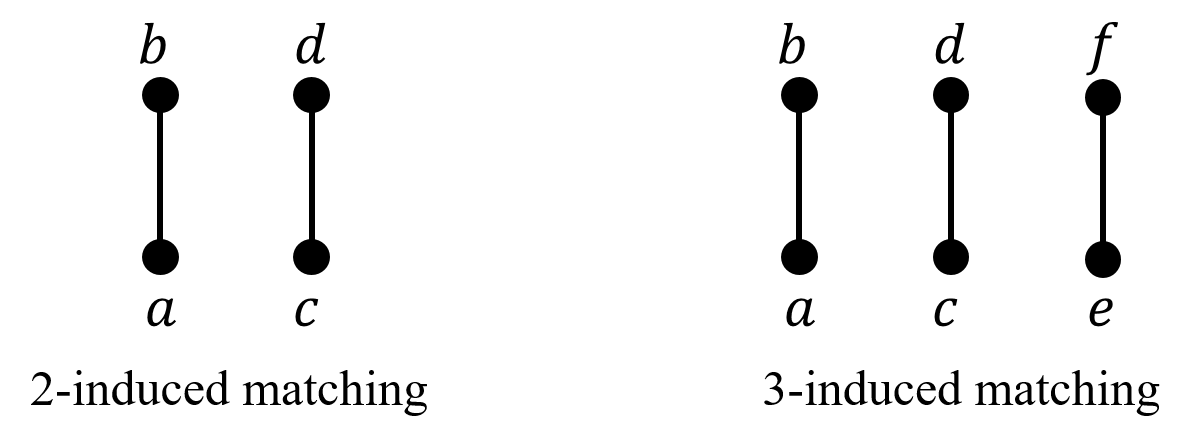}
		
		\caption{A $2$-induced matching and a $3$-induced matching.}
		
		\label{fig1}
	\end{center}
\end{figure}

A graph is called \emph{connected} if there is a path between every pair of its vertices. A graph  is called \emph{disconnected} if it is not connected. A \emph{connected component} of a graph $G$  is a maximal subgraph  where every  vertex is reachable from every other vertex in the same subgraph.  Therefore, every graph is partitioned into several connected components.
An \emph{elementary subgraph} is a subgraph in which the degree of each vertex is equal to $1$ or $2$  and each connected component of the subgraph is a regular graph. 
Let $\Lambda$ be an elementary subgraph of $G$ with  $c$ connected components and $s$ cycles. Then, $r(\Lambda)$ and $s(\Lambda)$ are defined as $r(\Lambda)=p(\Lambda)-c$ and $s(\Lambda)=s$, respectively.
We define $\mathcal{E}_{i}(G)$ as the set of elementary subgraphs of G that have $i$ vertices.
A subgraph $K$ of $G$ is  a \emph{vertex-induced subgraph}  if  it consists of a subset of vertices and all of the edges from $G$  that connect pairs of vertices in that subset.
Here, it is said that  $K$ is induced by $V(K)$. 
If $U \subseteq V(G)$, then $G-U$ is the subgraph induced by $V(G)-U$.
An induced matching is a set of independent edges whose endpoints induce no edges other than those in the set.
We call an induced matching with $i$ edges an $i$-induced matching. Fig. \ref{fig1} shows  a $2$-induced matching and a $3$-induced matching.

A Tanner graph is a bipartite graph with two sets of  \emph{variable nodes} and \emph{check nodes}.  
A Tanner graph  is called $(J,Z)$-regular or bi-regular if, for each variable node $v$, $d(v)=J$, and for each check node $w$, $d(w)=Z$.
If two graphs $G$ and $H$ are isomorphic, we write  $G\simeq H$.
If $G$ and $H$ are two graphs with $V(G)\cap V(H)=\phi$, then $G\cup H$ is a graph with the set of vertices $V(G)\cup V(H)$ and the set of edges $E(G)\cup E(H)$. 
Also, we denote the number of cycles of length $k$ in a graph $G$ by $M_{k}(G)$. The girth of a graph   is the minimum length of its cycles.
For a bipartite graph $G$ with girth $g$,   an approach for calculating $M_{k}(G)(k<2g)$ is to use the \emph{directed edge matrix} of $G$ \cite{Karimi 1}.   

\begin{figure}[h!]
	\begin{center}
		\includegraphics[scale=0.4]{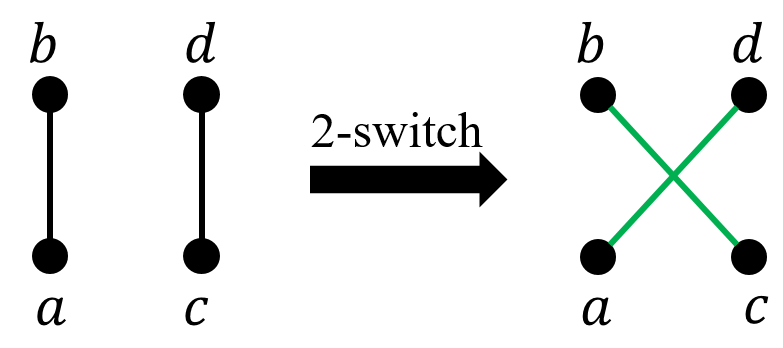}
		
		\caption{A 2-switch.}
		
		\label{fig2}
	\end{center}
\end{figure}

\begin{figure}[h!]
	\begin{center}
		\includegraphics[scale=0.372]{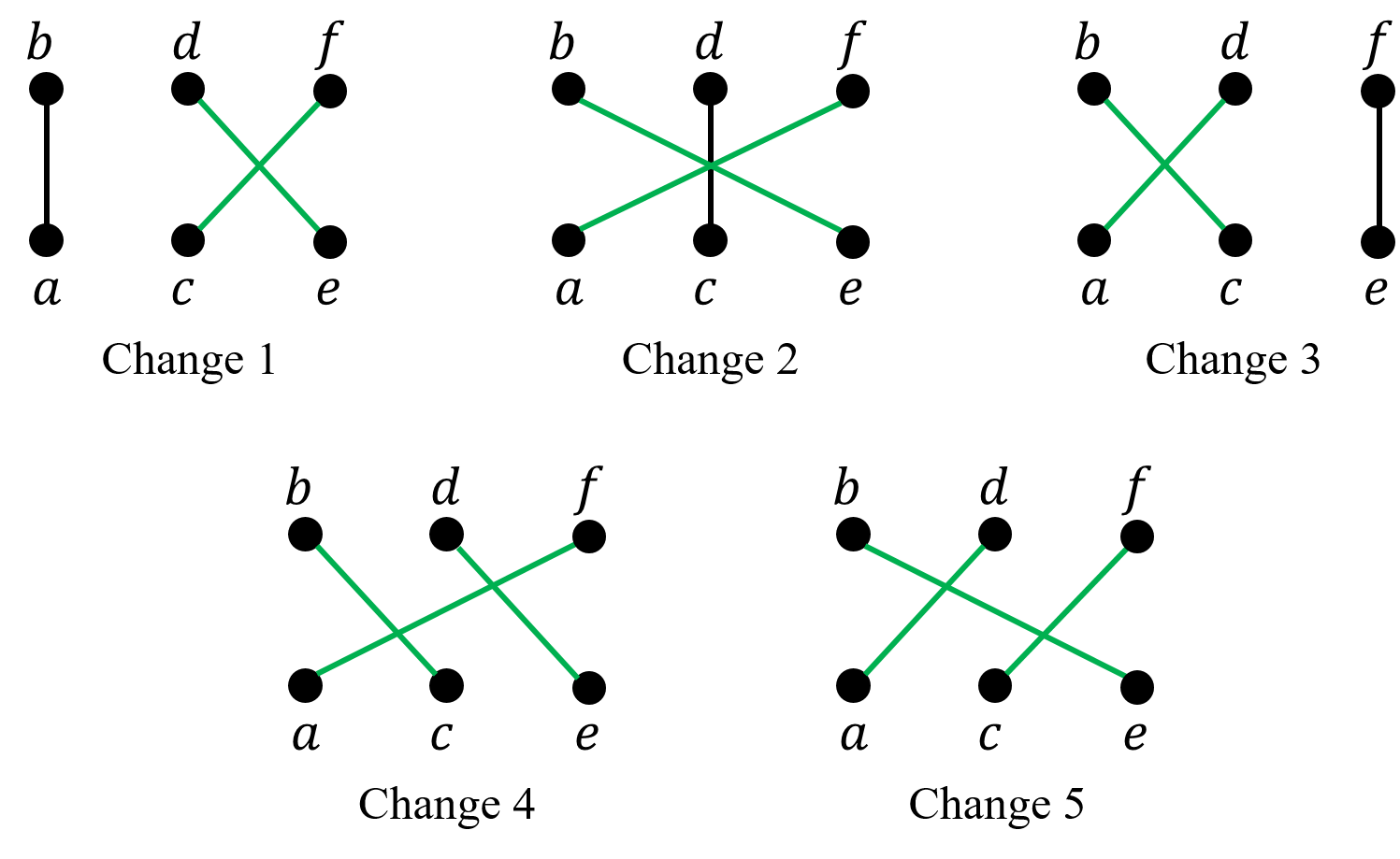}
		
		\caption{Five changes on a $3$-induced matching.}
		
		\label{fig3}
	\end{center}
\end{figure}

We denote  the characteristic polynomial of $G$ by  $f_{G}(x)$. Throughout the paper, the characteristic polynomial of a graph $G$ with $n$ vertices is considered as follows: 
\begin{equation}\label{equ1}
f_{G}(x)=x^{n}+c_1 x^{n-1}+c_2 x^{n-2} + \dots +c_{n-1}x +c_n.
\end{equation}
Also, we consider the eigenvalues of $G$ in the order $\lambda_{1}(G)\geq \dots \geq \lambda_{n}(G)$. The coefficients $c_1, \dots , c_n$ are related to the structure of $G$. 
A 2-switch operation is applied to a set of two independent edges such as $\{ab,cd\}$ with the condition $ad,bc \not \in E(G)$, as the edges $ab$ and $cd$ are removed and the edges $ad$ and $bc$ are added.  In Fig. \ref{fig2}, we can see a 2-switch operation. 
Provided that no edge is added to any part of a bipartite graph,
for a $2$-induced matching, we can apply only one 2-switch to the graph. For a $3$-induced matching, we consider five changes. In Fig. \ref{fig3},  these five changes are depicted. Clearly, each of these changes preserves the degree sequences of the graph, and the new graph is still bipartite.

\section{$3$-induced matchings in Tanner graphs}

First, we present a theorem  related to  the coefficients of the characteristic polynomial   and the elementary subgraphs.

\begin{Theorem}[{{\cite[Proposition 7.3]{Biggs 1}}}]\label{Theorem 1}
	Let $G$ be a graph with $n$ vertices and the characteristic polynomial in the form of (\ref{equ1}).
	Then, for every $1\leq i \leq n$, we have
	\begin{equation*}
	(-1)^{i}c_i=\sum_{\Lambda\in \mathcal{E}_{i}(G)}^{}(-1)^{r(\Lambda)}2^{s(\Lambda)}.
	\end{equation*}
\end{Theorem}

In the following, we provide a formula to enumerate the number of 3-induced matchings in a bipartite graph with a girth of at least $6$ (which results in providing a necessary and sufficient condition for the existence of a 3-induced matching).

\begin{Theorem}\label{Theorem 2}
	Suppose that $G$ is a bipartite graph with $n$ vertices, girth at least $6$  and the multiset of eigenvalues $\{\lambda_{r}\}_{r=1}^{n}$. Then,  $G$ has a $3$-induced matching if and only if 
	\begin{align*}\label{equ 17}
	\notag - \sum_{1\leq l_1 < l_2 < l_3 < l_4 < l_5 < l_6\leq n}^{}\lambda_{l_1}\lambda_{l_2}\lambda_{l_3}\lambda_{l_4}\lambda_{l_5}\lambda_{l_6}&\notag-\sum_{\stackrel{\{u,v,s,w\}}{uv,us,vw \in E(G)}}^{}(d(s)-1)(d(w)-1)\\&\notag -\sum_{\stackrel{ux, uy, uz \in E(G)}{x\not= y, y \not= z, x\not= z}}^{}\cfrac{(d(y)-1)(d(z)-1)}{2}\\&-\sum_{\stackrel{\{u,v,s,w\}}{uv,us,vw \in E(G)}}^{}q(G-(\{u,v,s,w\}\cup N(u,v,s,w)))\\&+2M_{6}(G)\geq 1\\ 
	\end{align*}

	\begin{proof}
		By Theorem \ref{Theorem 1}, for $i=6$,  we have
		\begin{equation*}
		(-1)^{6}c_6=\sum_{\Lambda\in \mathcal{E}_{6}(G)}^{}(-1)^{r(\Lambda)}2^{s(\Lambda)}.
		\end{equation*} 
		Since the girth of $G$ is at least $6$, it does not have any cycle  of length 4, and therefore 	$\mathcal{E}_{6}(G)$ consists of  cycles of length 6 (Type A) and three independent edges (Type B). If $\alpha$ and $\beta$ are the number of subgraphs of Type A and  Type B, respectively, then we have	
		\begin{equation*}
		c_6=(-1)^{6-1}\times 2^{1} \times \alpha +(-1)^{6-3}\times 2^{0} \times \beta= -2 \alpha-\beta.
		\end{equation*}
		We can consider five types for the elementary subgraphs of Type B.
		In Fig. \ref{fig4}, these five types have been shown ($ab$, $cd$ and $ef$ are three independent edges). 
		
		\begin{figure}[h!]
			\begin{center}
				\includegraphics[scale=0.382]{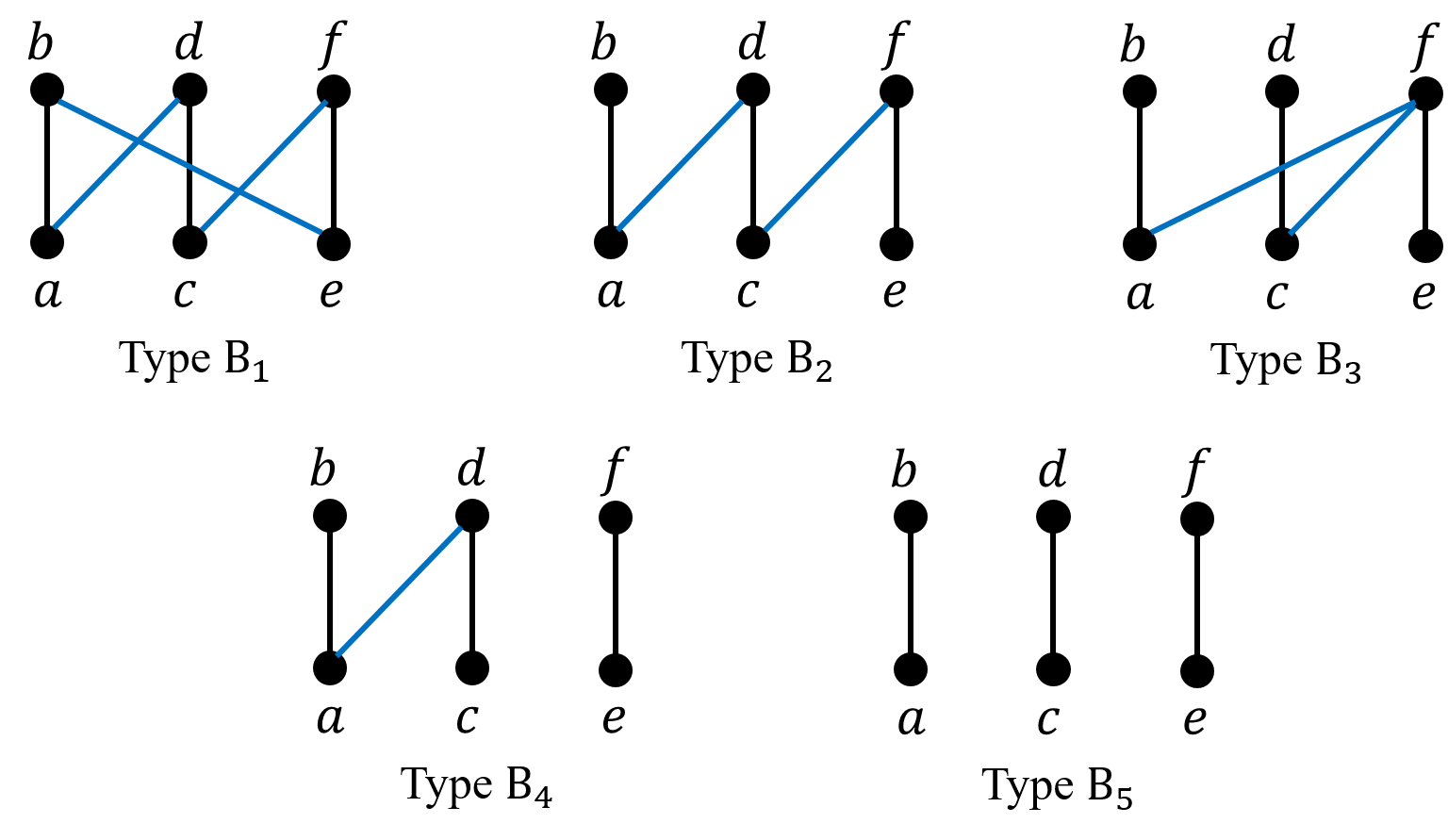}
				
				\caption{Five types of subgraphs of Type B.}
				
				\label{fig4}
			\end{center}
		\end{figure}
	\noindent
		We denote the number of subgraphs of Type $\mathrm{{B_1}}$, Type $\mathrm{{B_2}}$, Type $\mathrm{{B_3}}$, Type $\mathrm{{B_4}}$ and Type $\mathrm{{B_5}}$ by $\beta_1$, $\beta_2$, $\beta_3$, $\beta_4$ and $\beta_5$, respectively. 
		The values of $\beta_1$, $\beta_2$,  $\beta_3$ and $\beta_4$ are calculated as follows:
		\begin{equation*}
		\beta_1=2 \alpha, \ \beta_2=-6\alpha +\sum_{\stackrel{\{u,v,s,w\}}{uv,us,vw \in E(G)}}^{}(d(s)-1)(d(w)-1), \beta_3=\sum_{\stackrel{ux, uy, uz \in E(G)}{x\not= y, y \not= z, x\not= z}}^{}\cfrac{(d(y)-1)(d(z)-1)}{2},
		\end{equation*} 
		\begin{equation*}
		\beta_4=\sum_{\stackrel{\{u,v,s,w\}}{uv,us,vw \in E(G)}}^{}q(G-(\{u,v,s,w\}\cup N(u,v,s,w))).
		\end{equation*} 
		Note that for $\beta_3$, the division by 2 is because y and z can be interchanged in the summation index. Also, for $\beta_2$ and $\beta_4$, the summation is over all paths of length 3, and each path is considered exactly once. Now, we have
		\begin{align*}
		\notag c_6=-2\alpha-(2\alpha-6\alpha&+\sum_{\stackrel{\{u,v,s,w\}}{uv,us,vw \in E(G)}}^{}(d(s)-1)(d(w)-1)\\&\notag +\sum_{\stackrel{ux, uy, uz \in E(G)}{x\not= y, y \not= z, x\not= z}}^{}\cfrac{(d(y)-1)(d(z)-1)}{2}\\&\notag+\sum_{\stackrel{\{u,v,s,w\}}{uv,us,vw \in E(G)}}^{}q(G-(\{u,v,s,w\}\cup N(u,v,s,w)))\\&\notag+\beta_5).\\
		\end{align*}
		By definition of $M_{6}(G)$, we have  $\alpha=M_{6}(G)$.
		Now,   
		$
		c_6=\sum_{1\leq l_1 < l_2 < l_3 < l_4 < l_5 < l_6\leq n}^{}\lambda_{l_1}\lambda_{l_2}\lambda_{l_3}\lambda_{l_4}\lambda_{l_5}\lambda_{l_6}
		$
		\cite{Reichstein 1} implies that
		\begin{align*}
		\notag \beta_5=- \sum_{1\leq l_1 < l_2 < l_3 < l_4 < l_5 < l_6\leq n}^{}\lambda_{l_1}\lambda_{l_2}\lambda_{l_3}\lambda_{l_4}\lambda_{l_5}\lambda_{l_6}&\notag-\sum_{\stackrel{\{u,v,s,w\}}{uv,us,vw \in E(G)}}^{}(d(s)-1)(d(w)-1)\\&\notag -\sum_{\stackrel{ux, uy, uz \in E(G)}{x\not= y, y \not= z, x\not= z}}^{}\cfrac{(d(y)-1)(d(z)-1)}{2}\\&\notag-\sum_{\stackrel{\{u,v,s,w\}}{uv,us,vw \in E(G)}}^{}q(G-(\{u,v,s,w\}\cup N(u,v,s,w)))\\&+2M_{6}(G).\\ 
		\end{align*}
		Since $G$ has a $3$-induced matching if and only if $\beta_5\geq 1$, the proof is completed. 
		\hfill
	\end{proof}
\end{Theorem}

For calculating the number of 2-induced matchings in a bipartite graph with a similar approach, see \cite{Eshraghi 1}.
For 2-induced matchings,  $\mathcal{E}_{4}(G)$ has been considered.

\section{Numerical Results}\label{section 4}

In this section, we present some numerical results. We start by presenting an algorithm (\hspace{1sp}\cite{Eshraghi 1} is relevant).
The details of our algorithm are presented in Algorithm \ref{algo1}. We should   check the condition of Theorem \ref{Theorem 2}  for some steps $i$. This condition is to ensure the existence of a $3$-induced matching in the subgraph.   In step 2 of the algorithm, the set $K\subseteq \{6,\dots, n-1\}$ specifies the steps at which the condition of Theorem \ref{Theorem 2} must be checked. Which steps this condition is controlled in is an important issue. Further on in this section, we present some results about this issue.  
The conditions A, B, C and D in  the algorithm are presented in Table  \ref{tab1}. In fact, these are some methods based on the eigenvalues ​​of subgraphs for moving from an initial graph toward its six vertices.	In step 3, when we find $G_{6}\simeq K_{2}\cup K_{2} \cup K_{2}$, we  apply  five changes separately, as depicted in Fig. \ref{fig3}, and according to  step 3 of the algorithm, the work continues. 
\renewcommand{\algorithmicrequire}{\textbf{Input:}}
\renewcommand{\algorithmicensure}{\textbf{Output:}}

\begin{algorithm}[h!]
	\caption{}\label{algo1}
	\begin{algorithmic}
		\Require A bipartite graph $G=G_{in}$ with $n$ vertices ($n\geq 7$) and girth  $6$,  a set $K\subseteq \{6,\dots, n-1\}$ and a condition $\text{R}\in \{\text{A,B,C,D}\}$ (Table \ref{tab1}).
		\Ensure Graph $G_{out}$, the last graph $H$ with the minimum value of $M_{6}(H)$ during $n$ iterations. 
		\State  \textbf{Step 1.} Set $G_n=G$.
		\State \textbf{Step 2.} For $i=n-1$ to $6$, let $\Gamma$ be the set of all vertex-induced subgraphs $S$ with $p(S)=i$  such that $V(S)\subset V(G_{i+1})$,  and if $i \in K$, then $S$ has at least one 3-induced matching (according to Theorem \ref{Theorem 2}).
		If $\Gamma=\phi$, then end the process and declare that an error occurred. Otherwise, set $G_i$ to be a subgraph that satisfies the condition R  among all elements of $\Gamma$. If  there is more than one subgraph that has this property, then
		randomly choose  a subgraph that  has this property.
		\State \textbf{Step 3.} If $G_6\not \simeq K_2\cup K_2 \cup K_2$, then end the process and declare that an error occurred. If $G_6 \simeq K_2\cup K_2 \cup K_2$, then apply 
		the five changes introduced in Fig. \ref{fig3} separately. Among the five new graphs, select the graph with girth  at least 6 that has the least number of cycles of length 6. If there is more than one graph with these properties,  select a graph constructed with these properties at random. Consider the resulting graph $H$. If all five new graphs have girth $4$, set $H=G$ (no change occurs).
		\State \textbf{Step 4.} If $M_{6}(H)\geq 1$ and  the number of iterations is less than or equal to  $n$, then set $G=H$ and go to step 1. 
	\end{algorithmic}
\end{algorithm}

\begin{table}[h!]
	
	\begin{center}
		\caption{Conditions A, B, C and D in  Algorithm \ref{algo1}}

		\begin{tabular}{|c|c|c|}
			
			\hline
			Condition	&     Minimum/Maximum  &  The Value     \\  
			\hline
			A	&    Min  & $\lambda_{1}(S)-\lambda_{2}(S)$     \\  
			\hline
			B	&    Min  & $\lambda_{1}(S)-\lambda_{3}(S)$     \\  
			\hline
			C	&    Max  & $\lambda_{1}(S)+\lambda_{2}(S)$     \\  
			\hline
			D	&    Max  & $\lambda_{1}(S)+\lambda_{2}(S)+\lambda_{3}(S)$     \\  
			\hline
			
		\end{tabular}
		
		\label{tab1}
		
	\end{center}

\end{table}

\begin{figure}[h!]
	
	\begin{center}
		\includegraphics[scale=0.578]{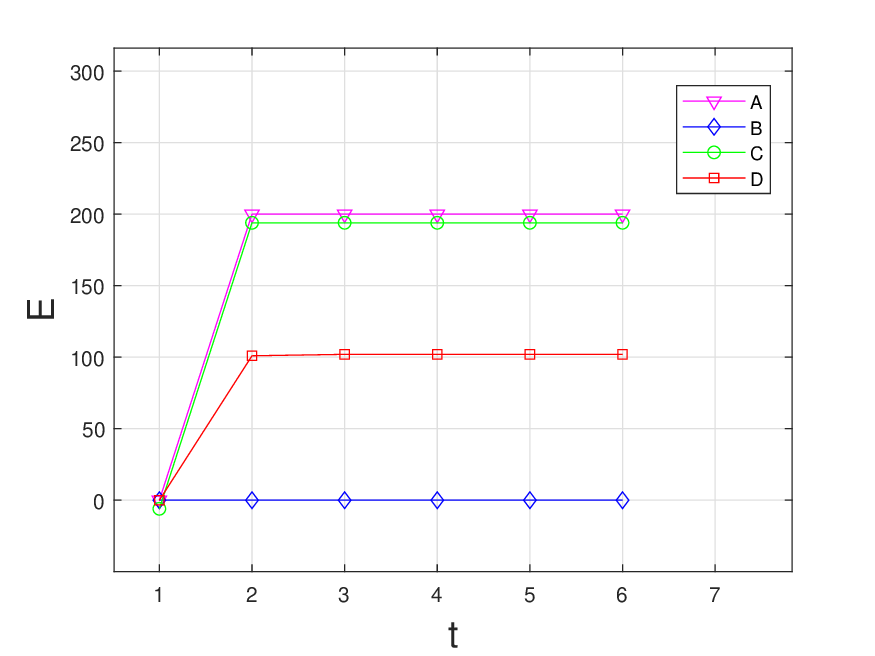}
	\end{center}
	\vspace{-0.55cm}
	\caption{The number of errors in 200 selected $(3,6)$-regular graphs with girth $6$ for some values of $t$.}
	
	\label{fig5}
\end{figure}

\begin{figure}[h!]
	
	\begin{center}
		\includegraphics[scale=0.578]{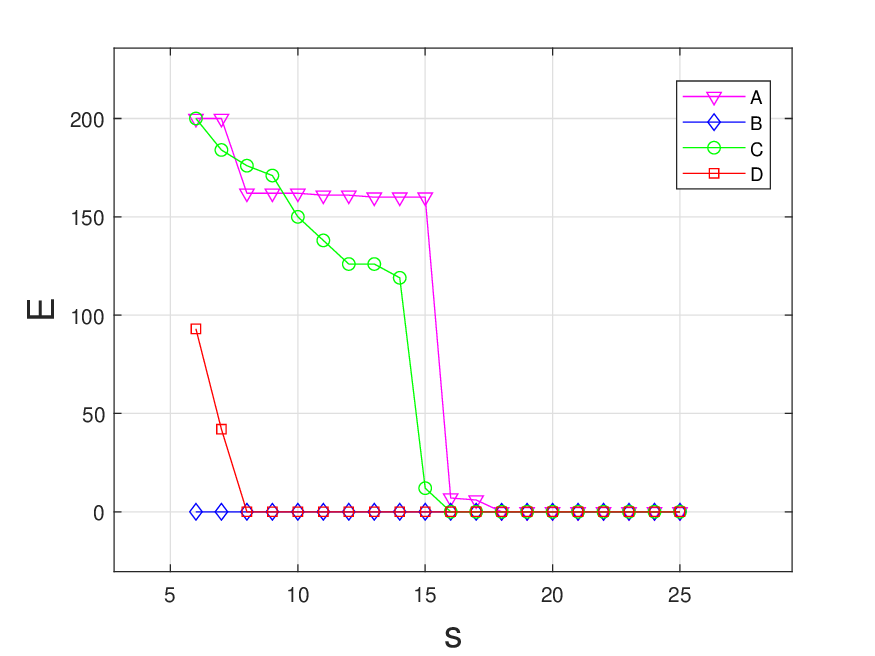}
	\end{center}
	\vspace{-0.55cm}
	\caption{The number of errors in 200 selected $(3,6)$-regular graphs with girth $6$ for some values of $s$.}
	
	\label{fig6}
\end{figure}

Now, we select 200 $(3,6)$-regular graphs with girth $6$ and $n$ vertices ($54\leq n \leq 270$). The selected graphs are entered into the process, and the algorithm runs for one iteration. 
We consider $K\subseteq \{6,\dots , 24\}$ and set $t$ as the distance between steps (by starting with $6$)  for which the condition of Theorem \ref{Theorem 2} is checked.  For example, $t=2$ means the steps $6,8,10, \dots , 24$.  In Fig. \ref{fig5}, we see the number of errors $(E)$ in the selected graphs for some values of $t$ (for condition C, the curve has been plotted slightly lower to illustrate the distinction. In fact, the curves of A and C  coincide). An error means that a 3-induced matching is not obtained. As can be seen, the distance between  steps should be very small for the conditions A, C and D. Distance one seems the best selection.
Now, we again  consider 200 $(3,6)$-regular graphs  with girth $6$ and $n$ vertices ($54\leq n \leq 270$). We set $K=\{6,\dots,s\}$  (the distance  between steps is one). In Fig. \ref{fig6}, we see the results for different values of $s$.  As we see, when $\lambda_{3}$ is involved (conditions B and D), for a larger number of values ​of s, the number of errors is zero.
Note that as the graph becomes denser than $(3,6)$-regular, we generally expect the number of errors to increase.

In Table \ref{tab2}, we see the  results for the number of cycles of length 6 for some bi-regular input graphs with girth $6$,  randomly generated. Here,
$M_{6}(G_{in})$ is the number of cycles of length $6$ in the input graph, and $M_{6}(G_{out}^{A})$,  $M_{6}(G_{out}^{B})$, $M_{6}(G_{out}^{C})$ and $M_{6}(G_{out}^{D})$ are the number of cycles of length $6$ in output graphs with conditions A, B, C and D, respectively.
The condition of Theorem \ref{Theorem 2} has been used for $i\leq30$
.
As can be seen, in all four  types of conditions, using 3-induced matchings can be effective for reducing the number of cycles of length $6$. For some initial graphs in Table \ref{tab2},   conditions  A and C are more effective than   conditions  B and D.

\begin{table}[h!]
	
	\begin{center}
		\caption{The number of cycles of length $6$ in output graphs obtained by  the conditions  in Table \ref{tab1}
			for some bi-regular graphs}

			\begin{tabular}{|c|c|c|c|c|c|c|c|c|c|c|c|}
				
				\hline
				$n$	&     $J$  &  $Z$   & $M_{6}(G_{in})$  &  $M_{6}(G_{out}^{A})$  &   $M_{6}(G_{out}^{B})$ &  $M_{6}(G_{out}^{C})$ &  $M_{6}(G_{out}^{D})$
				\\  
				\hline
				
				$210$	&  $3$  &  $3$   &  $16$  &  $0$  &   $0$  &  $9$ &  $0$   \\
				\hline
				
				$224$	&  $4$  &  $4$    &  $141$   &   $24$   &   $26$  & $47$  &  $49$  \\
				\hline
				
				$133$	&  $3$  &  $4$    &  $34$   &   $0$   &   $1$ & $4$ &  $4$  \\
				\hline
				
				$217$	&  $3$  &  $4$    &  $51$   &   $0$   &   $0$ &  $2$ &  $0$   \\
				\hline
				
				$160$	&  $3$  &  $5$    &  $101$   &   $19$   &   $32$  &  $31$ &  $41$   \\
				\hline
				
				$240$	&  $3$  &  $5$    &  $84$   &   $5$   &   $12$ &  $16$  &  $17$   \\
				\hline
				
				$160$	&  $2$  &  $6$    &  $25$   &   $0$   &   $0$ & $0$ &  $0$   \\
				\hline
				
				$180$	&  $3$  &  $6$    &  $179$   &   $89$   &   $119$ & $98$ &  $94$ \\
				\hline
				
				$216$	&  $3$  &  $6$    &  $186$   &   $75$   &   $99$ & $83$ &  $90$ \\
				\hline
				
				$270$	&  $3$  &  $6$    &  $152$   &   $59$   &   $73$  & $57$ &  $61$   \\
				\hline

			\end{tabular}

		\label{tab2}
		
	\end{center}

\end{table} 

\newpage

\section{Conclusion}

In this paper, we investigated using 3-induced matchings and eigenvalues of subgraphs to apply some changes to Tanner graphs. We presented a formula for the number of 3-induced matchings in bipartite graphs with girth at least $6$. Also,  we presented a process in which it was possible to select from four types of strategies based on the eigenvalues ​​of subgraphs. We provided some  results regarding the application of the condition of Theorem \ref{Theorem 2} during the process, as well as the effect of the presented process on the number of cycles of length 6 in some graphs.

\section*{Acknowledgments}

This work is based upon research funded by Iran National Science Foundation (INSF) under project No. 4035789 and 
the second author was in part supported by a grant from IPM (No. $1405130118$).

\end{document}